\documentclass[10pt]{article}

\usepackage[margin=1in]{geometry}
\usepackage[utf8]{inputenc}
\usepackage[T1]{fontenc}
\usepackage[english]{babel}
\usepackage{graphicx}
\usepackage{amsmath,amssymb}
\usepackage{booktabs}
\usepackage{array}
\usepackage{longtable}
\usepackage{xcolor}
\usepackage{caption}
\usepackage{placeins}
\usepackage[numbers,sort&compress]{natbib}
\usepackage{xr-hyper}
\usepackage[colorlinks=true,linkcolor=blue,citecolor=blue,urlcolor=blue]{hyperref}
\usepackage{lineno}
\usepackage{ulem}

\title{Ordered-to-disordered transfer learning with graph neural networks for formation-energy and HOMO-LUMO gap prediction in high-entropy perovskite oxides}

\author{
Panupol Untarabut$^{1}$,
Narjes Jomaa$^{1}$,
Sylvian Cadars$^{1}$,
Olivier Masson$^{1}$,\\
Samuel Bernard$^{1}$,
Assil Bouzid$^{1,\ast}$,
Santanu Saha$^{1,\ast}$
}

\date{}

\begin{document}
\maketitle

\noindent
$^{1}$University of Limoges, CNRS, IRCER, UMR 7315, F-87000, Limoges, France\\

\noindent
$^{\ast}$Correspondence: assil.bouzid@cnrs.fr; santanu.saha@unilim.fr

\begin{abstract}
High-entropy perovskite oxides (HEPOs) represent a chemically complex class of materials with promising functional properties, yet their vast compositional space and chemical/structural disorder pose significant challenge for accurate property prediction. Graph neural networks (GNNs) enable rapid exploration of materials space but are often limited by the availability of representative training data. Here, we investigate ordered-to-disordered transfer learning using GNNs for formation-energy and HOMO–LUMO gap prediction in HEPOs by transferring knowledge learned from chemically ordered perovskites. Four representative GNN models, including CGCNN, GATGNN, ALIGNN and M3GNet are evaluated to understand the role of structural representations, spanning pairwise two-body and angular three-body interactions in transfer performance. We find strong property-dependent transfer behavior: formation-energy prediction transfers effectively to disordered HEPOs, whereas HOMO–LUMO gap prediction shows limited transferability due to its sensitivity to local chemical environments. Incorporating a small HEPO-specific training dataset substantially improves HOMO–LUMO gap prediction. Representation-level analysis using UMAP further highlights the importance of encoding three-body geometric information such as in ALIGNN for capturing complex structure–property relationships and improving transferability.
\end{abstract}

\section*{Introduction}

High-entropy oxides extend the concept of configurational-entropy stabilization to ceramic materials by mixing several cation species across one or more crystallographic sites~\cite{rost2015entropy}. This concept has been successfully extended to the perovskite structural family, giving rise to high-entropy perovskite oxides (HEPOs)
~\cite{zhang2019review}. In ABO$_3$ perovskite oxides, this chemical flexibility enables the incorporation of many different A-site and/or B-site cations while maintaining a connected network of corner-sharing BO$_6$ octahedra~\cite{jiang2018new,sarkar2018rare}. Similarly, the family of ordered double perovskites is also found to support a broad range of stability, electronic, and optical properties~\cite{shah2024first}.
As for HEPOs, they provide a large chemical space for controlling phase stability, electronic structure, ionic transport, catalytic activity, and other functional properties~\cite{schweidler2024high}.
Their phase stability also depends on synthesis conditions, particularly the thermal cycle and oxygen partial pressure, which determine the preferred cation oxidation  states; thereby promoting the formation of a single-phase high-entropy oxide~\cite{almishal2025thermodynamics}.
However, the chemical disorder that gives HEPOs their useful properties also makes them difficult to investigate, particularly through computational modeling.
Even within a macroscopically homogeneous single-phase HEPO, B-site cation disorder produces a distribution of atomic-scale cation environments, local strains, and BO4$_6$ octahedral distortions~\cite{su2022direct}. These local variations can influence the charge distribution and leads to a broad distribution of electronic properties~\cite{liang2021electronic,bai2024high}.

First-principles calculations based on density functional theory (DFT)~\cite{hohenberg1964inhomogeneous,kohn1965self} provide a reliable framework for investigating the stability and electronic properties of HEPOs.
However, modeling chemical disorder with DFT is computationally demanding because it requires large supercells, extensive structural relaxations, and often multiple representative cation configurations to capture the diversity of local atomic environments.
This high cost motivates the use of machine-learning models that learn relationships between atomic structures and properties from a limited set of DFT calculations and then extrapolate to a much larger chemical and configurational space~\cite{emery2017high,bare2023dataset,kim2024machine}.
For example, machine learning models can predict synthesizability of a doped perovskite from limited experimental data, while transfer-learning and active-learning strategies can accelerate the discovery of multimetallic perovskite catalysts~\cite{zhai2022predicting,jiang2024transfer,moon2024active}.

A particularly promising direction involves developing machine learning models that are able to extend insights from simpler, ordered perovskite oxides and transfer the learning to the more complex and chemically disordered high-entropy perovskite oxides. But, these HEPO structures often contain a vast diversity of local chemical environments, a large portion of which are often not explicitly represented in the training data, posing a major challenge for property prediction~\cite{li2023center,gupta2024structure}.

Ordered perovskite oxides offer an ideal starting point for exploring this transferability. Simple  perovskites (ABO$_3$) and ordered double perovskites (A$_2$BB'O$_6$) share fundamental structural and chemical features with HEPOs: they comprise the same set of A-site and B-site elements, exhibit identical oxygen coordination environments, and maintain the characteristic BO$_6$ octahedral networks. However, they lack the full complexity of multi-element A-site/B-site disorder, which is a hallmark of HEPOs~\cite{vasala2015a2b,chen2019rare}.

Crucially, the stability and electronic structure of these materials depend on the local cation environment and local distortions which is manifested via the associated bond lengths, bond angles, and octahedral tilting. Models trained on ordered perovskite oxides may therefore capture transferable knowledge --- such as B--O bond lengths, B--O--B angles, octahedral tilting, and the chemistry of A-site and B-site elements --- providing a robust foundation for understanding and predicting the behavior of more disordered systems~\cite{glazer1972classification,woodward1997octahedral,ray2017effects,ghosh2022insights}.

Among different available machine-learning models, graph neural network (GNN) are suitable 
for this task as their graphical representation of crystal structures enables direct incorporation of local chemical environment~\cite{CGCNN}. Various GNN models relying on different architectures have been reported.
Among them, Crystal Graph Convolutional Neural Network (CGCNN) represents crystalline materials as atom–bond graphs, where atoms are nodes and inter-atomic connections encode the local coordination environment~\cite{CGCNN}. Graph Attention-based Graph Neural Network (GATGNN) extends graph-based crystal modeling with attention mechanisms that learn the relative importance of neighboring atoms and graph-level features~\cite{GATGNN}. Atomistic Line Graph Neural Network (ALIGNN) incorporates both bond and bond-angle information by jointly operating on the atomic graph and its corresponding line graph~\cite{ALIGNN}, while Materials Graph Neural Network (M3GNet) captures higher-order interactions through geometric three-body features and message passing, enabling the modeling of complex many-body atomic environments~\cite{M3GNet}.

These architectural differences are especially important for perovskite oxides, as bond angles and octahedral tilting can strongly influence both stability and electronic structure~\cite{woodward1997octahedral,ray2017effects,su2022direct}. Comparing these different models by using the same ordered-to-disordered transfer test can therefore reveal which structural representations and model architectures are most effective for predicting the properties of HEPOs.

In this work, we construct a DFT dataset containing ordered single ABO$_3$ \cite{untarabut2026single}, double A$_2$BB'O$_6$ perovskites~\cite{untarabut2026double} and special quasirandom structure (SQS)-generated HEPO models~\cite{jomaa2026aiida,untarabut2026entropy} based on the same combination of A-site and B-site elements~\cite{SQS}. The SQS approach provides an efficient representation of random chemical disorder 
while keeping the supercell size computationally tractable. We train CGCNN, GATGNN, ALIGNN, and M3GNet models to predict formation energies calculated relative to the corresponding binary oxides ($E_f$) and DFT HOMO-LUMO gaps ($E_g$) as measures of thermodynamic stability and electronic structure, respectively.
We first evaluate the models on held-out ordered perovskites and then test their transferability to HEPO structures. We also analyze errors for different chemical groups, the features of the graph representations learned by the models, and the minimal amount of HEPO-specific training data required to improve the ordered-domain models. Models trained only on ordered perovskites show stronger transferability for formation energy than for HOMO-LUMO gap prediction, whereas incorporating a limited fraction of HEPO data substantially improves predictions in the chemically disordered domain. In this context, the ordered domain encompasses the ordered single- and double-perovskite datasets, whereas the disordered domain corresponds to the HEPO dataset.

\section*{Results}

\subsection*{Dataset construction}

Different crystal structures models and their computed DFT properties used in this study have been generated in our previous works on simple and double perovskite oxides. The ordered domain contained simple ABO$_3$ perovskites and ordered double-perovskite A$_2$BB'O$_6$ structures~\cite{untarabut2026single,untarabut2026double}. Within this study, the A-site cations are restricted to [Ca, Sr, Ba] and B-site cations restricted to [Ti, Zr, Hf, Sn, Ge] chemical species.
This elemental space produced 15 simple ABO$_3$ compositions and 30 distinct A$_2$BB'O$_6$ compositions. Each composition included multiple crystal templates, B/B$'$ orderings, and octahedral tilting patterns, as described in our previous works~\cite{untarabut2026double}.  After DFT relaxation and filtering, the ordered-domain dataset contained 10898 structures, including 3024 ABO$_3$ structures and 7874 A$_2$BB'O$_6$ structures.~\cite{untarabut2026single,untarabut2026double}

The HEPO domain was constructed separately using different combination of supercell shape, size and SQS models to provide a chemically disordered transfer domain. For each A-site family, the five B-site elements were combined in an equimolar concentration within a special quasirandom structure (SQS), resulting  in a nominal composition A$_5$TiZrHfSnGeO$_{15}$, where A is Ca, Sr, or Ba~\cite{SQS}.
This procedure generated Ca-, Sr-, and Ba-based HEPO families comprising 1810 relaxed structures in total ~\cite{jomaa2026aiida,untarabut2026entropy}. As the ordered and HEPO domains share the same set of A-site and B-site elemental pools, their comparison isolates the effect of local chemical disorder from that of introducing new chemical elements.

For each structure in the dataset, DFT~\cite{hohenberg1964inhomogeneous,kohn1965self} calculation was carried out to obtain the binary oxide-referenced formation energy ($E_f$) and electronic HOMO-LUMO gap ($E_g$), which served as the two machine-learning model targets.
The ordered dataset was divided into training, validation, and test subsets using a 
a 80:10:10 split for each perovskite composition, 
corresponding to 8722, 1088, and 1088 structures, respectively.
Among them, the training subset was used to optimize the model parameters, the validation subset to monitor performance and apply early stopping, and finally the test subset was reserved for the final ordered-domain evaluation.

Further, the HEPO dataset was independently divided within each A-site family using the same 80:10:10 ratio, giving 1448 training, 181 validation, and 181 test structures. Two training strategies were considered here:
(a) First, blind transfer test, where models were trained and validated only on the ordered datasets, and then the model's performance was evaluated directly on the HEPO test subset and (b) Secondly, a fine-tuning and data-efficiency strategy was applied where fractions of 0.0, 0.2, 0.4, 0.6, 0.8, and 1.0 of the HEPO training subset were combined with the ordered training data to create a new training set.
The HEPO validation subset was then used for model selection, while the HEPO test subset remained fixed and was used only for final evaluation.
Composition-resolved counts and the complete split assignments are provided in Table~S1 of Section~S1 in the SI.

Hence, this strategic design of our perovskite dataset allows us to address three questions:
\begin{itemize}
    \item How accurately can the trained GNN models interpolate within the ordered-perovskite domain?
    \item Are the structure-property relationships learned by the GNN models from ordered perovskites transferable directly to HEPOs?
    \item How much HEPO-specific training data are needed to achieve accurate and transferable predictions?
\end{itemize}

\subsection*{Training models on ordered domain}

\begin{figure}[!htbp]
\centering
\includegraphics[width=0.45\textwidth]{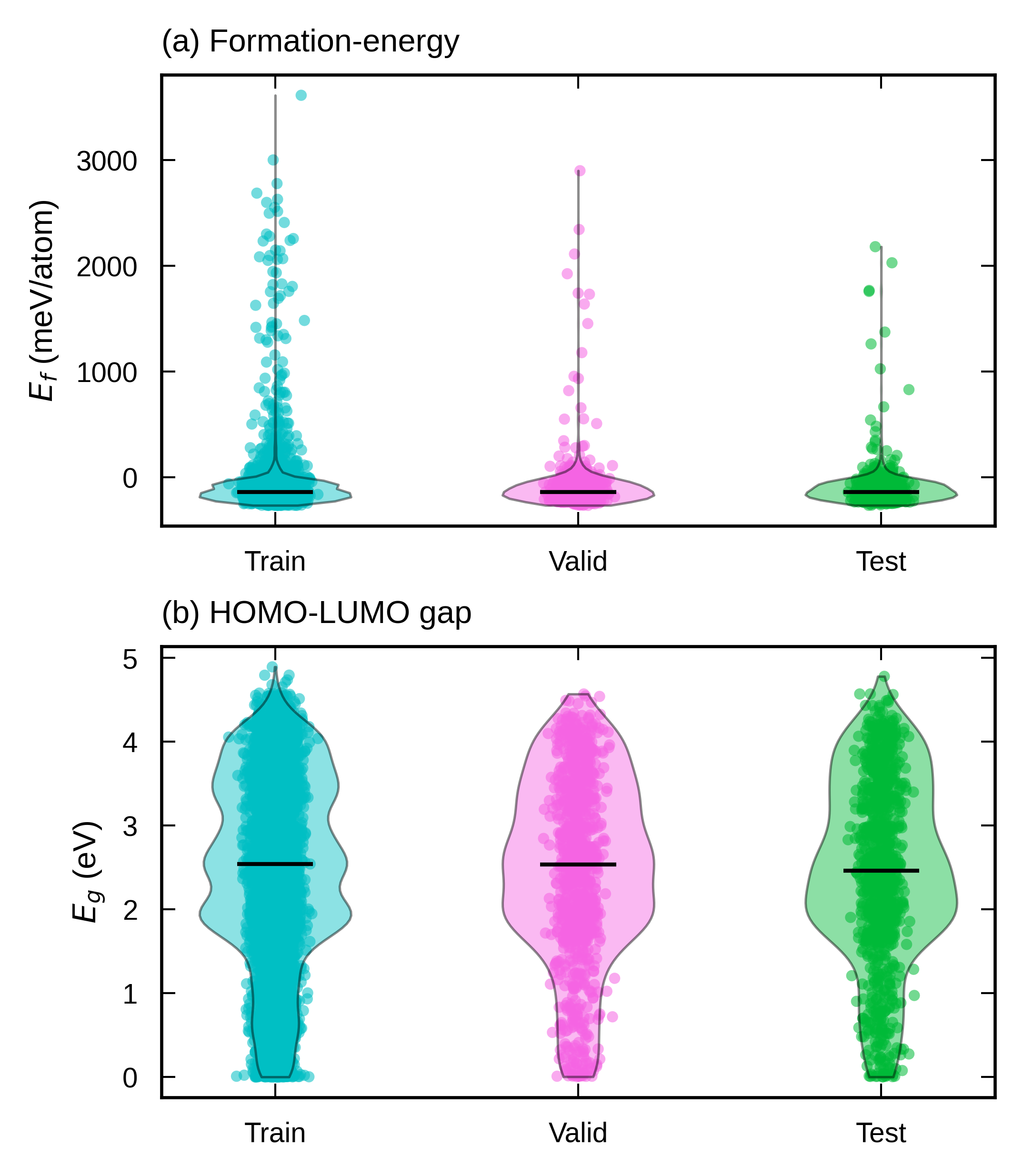}
\caption{Target-property distributions for the training, validation, and test subsets of the ordered single and double perovskite oxide dataset: (a) formation energy (meV/atom) and (b) HOMO-LUMO gap (eV). Formation energy is evaluated with respect to stable binary oxides of different cationic species. Ordered perovskite oxide consists of single ABO$_3$ and double A$_2$BB'O$_6$ perovskites.}
\label{fig01:dataset_split}
\end{figure}

Graph neural network architectures were trained as scalar regressors for two DFT target properties: the binary oxide-referenced formation energy ($E_f$) and 
the electronic HOMO-LUMO gap ($E_g$).
The definition of formation-energy and electronic-structure calculation settings are described in the Sec.\textit{Methods}, with the reference phases listed in Table~S2 of Section~S2 in the SI.
The target distributions for the ordered-perovskite training, validation, and test subsets are shown in Fig.~\ref{fig01:dataset_split}.
These three subsets cover similar ranges for both target properties, indicating that the composition-stratified 80:10:10 split provides a consistent interpolation benchmark.
The HOMO-LUMO gap distribution spans both low-gap and wide-gap structural models, while the formation-energy distribution contains a dense low-energy region and a rapidly decaying tail in the higher-energy region.
This diversity in their properties is useful for evaluating the performance of the trained GNN models both across the 
dominant trends and the less common high-error regions of the explored sample space.

\FloatBarrier

Four graph neural network architectures, CGCNN~\cite{CGCNN}, GATGNN~\cite{GATGNN}, ALIGNN~\cite{ALIGNN}, and M3GNet~\cite{M3GNet},
were trained and evaluated using the same training, validation, and test subsets.
This ensured that all models were compared using identical structural data splits.
All models were trained to predict the target properties using normalized target values computed from the training set.
The optimized objective was the mean squared error (MSE) between the predicted and DFT reference values after normalization:

\begin{equation}
\mathcal{L}_{\mathrm{MSE}} =
\frac{1}{N}\sum_{i=1}^{N}
\left(\hat{y}^{\,\mathrm{norm}}_i - y^{\mathrm{norm}}_i\right)^{2},
\end{equation}

where $N$ is the number of structures, $\hat{y}^{\,\mathrm{norm}}_i$ is the model prediction in normalized target space, and $y^{\mathrm{norm}}_i$ is the normalized DFT value.
The normalized target is defined as:

\begin{equation}
y^{\mathrm{norm}}_i =
\frac{y_i - \mu_{\mathrm{train}}}{\sigma_{\mathrm{train}}},
\end{equation}
where $\mu_{\mathrm{train}}$ and $\sigma_{\mathrm{train}}$ are the mean and standard deviation of the target values in the training set, respectively.

Adam optimization~\cite{kingma2014adam} was used for all models with a learning rate of 0.001 and a weight decay of $1\times10^{-5}$.
A multi-step learning-rate scheduler reduced the learning rate at epochs 50, 100, 200, 400, and 700 (see Section.~S3 in the SI for more details).
Training continued until no improvement in the validation loss was observed for 100 consecutive epochs, at which point early stopping was applied. Detailed descriptions of the model architectures and their hyper-parameters are provided in Table~S3 in Section~S3 of the SI.
For each model, the checkpoint corresponding to the minimum validation loss was selected as the best-performing model.

\begin{figure}[!htbp]
\centering
\includegraphics[width=0.95\textwidth]{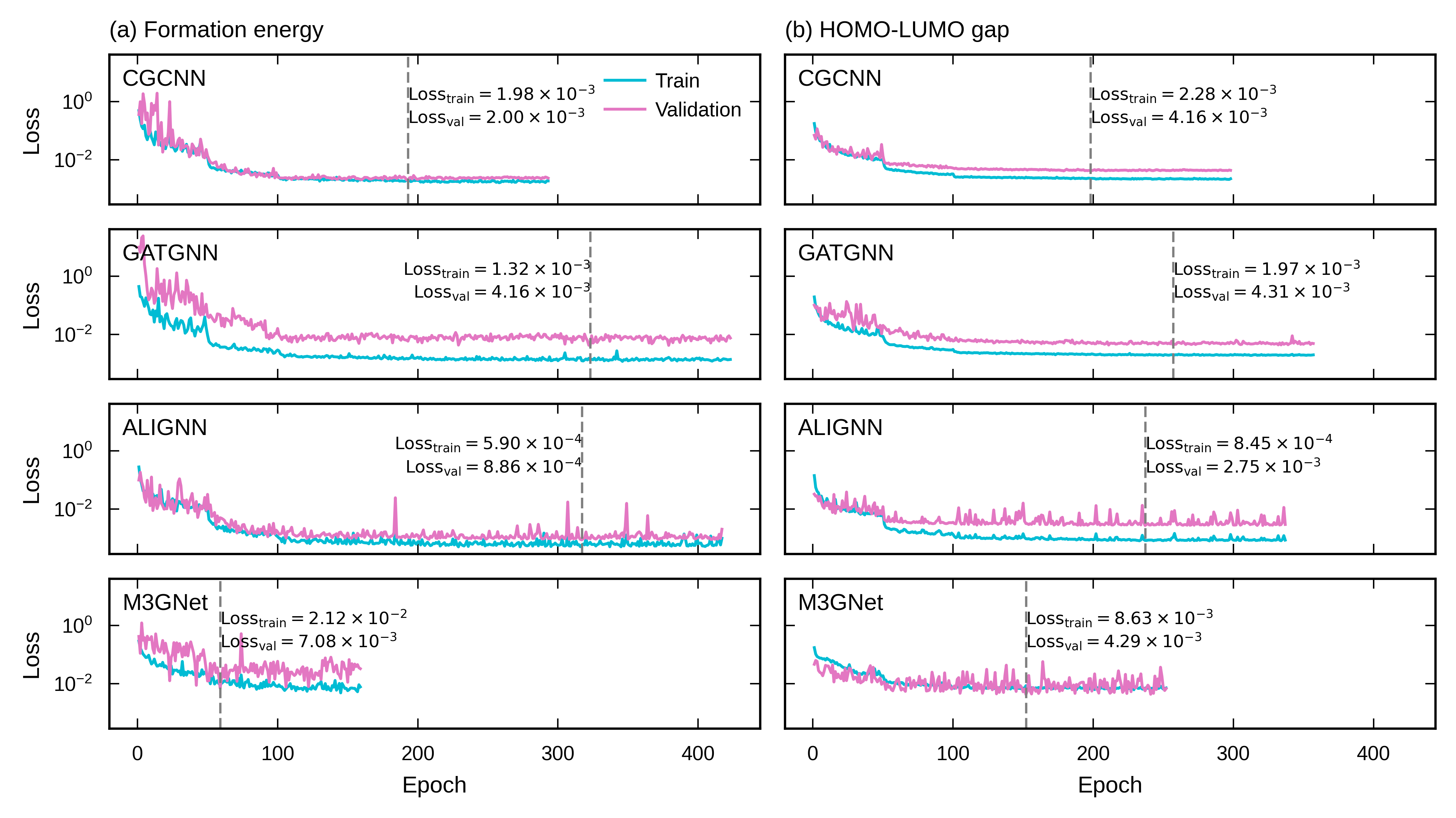}
\caption{Training behavior of the graph neural network models on the training and validation set of the ordered single and double perovskite dataset. (a) Formation-energy($E_f$) training and validation loss trajectories for CGCNN, GATGNN, ALIGNN, and M3GNet. (b) HOMO-LUMO gap ($E_g$) training and validation loss trajectories for the same models. The cyan and pink solid lines represent the training and validation losses, respectively. Grey vertical dashed lines mark the best validation checkpoints.}
\label{fig02:training-history}
\end{figure}

Training histories of these GNN models are shown in Fig.~\ref{fig02:training-history}. All models show a rapid decrease in loss during the early epochs, followed by slower convergence.
This indicates that the models quickly learn the dominant structure--property relationships in the ordered perovskite dataset and then progressively refine their predictions during later training.
The final model was selected from the epoch with the minimum validation loss, as marked by the grey vertical dashed line in Fig.~\ref{fig02:training-history}.

\FloatBarrier

For formation-energy prediction (Fig.~\ref{fig02:training-history}a), all models show a rapid early loss reduction followed by slower refinement.
CGCNN converges to nearly identical training and validation losses of $1.98\times10^{-3}$ and $2.00\times10^{-3}$, respectively, indicating consistent performance across both subsets. GATGNN reaches a lower training loss of $1.32\times10^{-3}$, while its validation loss remains higher at $4.16\times10^{-3}$, resulting in a larger training--validation gap than that observed for CGCNN.
ALIGNN gives the best formation-energy convergence, with the lowest training and validation losses, i.e. $5.90\times10^{-4}$ and $8.86\times10^{-4}$, respectively.
This improvement is consistent with ALIGNN's use of both the atomic graph and its line graph, in which bonds are represented as nodes and connections between adjacent bonds encode bond-angle information. This explicit angular representation is relevant to perovskite tilting and octahedral distortions~\cite{glazer1972classification,woodward1997octahedral,xiang2017rules}. On contrary,
M3GNet shows the weakest convergence for this scalar-regression setup, with training and validation losses of $2.12\times10^{-2}$ and $7.08\times10^{-3}$, respectively.
Although M3GNet was originally designed for energy, force, and stress learning with explicit many-body information~\cite{M3GNet}, the present configuration is a single-property regression model and is not optimized as a full interatomic potential.

For HOMO-LUMO gap prediction (Fig.~\ref{fig02:training-history}b), ALIGNN again achieves the lowest validation loss of ($2.75\times10^{-3}$), while CGCNN, GATGNN, and M3GNet converge to validation losses of ($4.16\times10^{-3}$), ($4.31\times10^{-3}$), and ($4.29\times10^{-3}$), respectively. ALIGNN explicitly incorporates angles between adjacent bonds through message passing on the atomic graph and its line graph. Its lower validation loss is therefore consistent with the importance of B--O--B angles and octahedral distortions in controlling orbital overlap and the HOMO-LUMO gap of perovskite oxides~\cite{ray2017effects}.

Overall, the training histories identify ALIGNN as the strongest architecture for both formation-energy and HOMO-LUMO gap properties and CGCNN as a stable baseline. In addition, these results highlight that the inclusion of explicit angular information in the model architecture improves convergence for this perovskite dataset. We now focus on the final model ranking on the held-out test set.

\subsection*{Prediction performance on ordered domain}
\begin{figure}[!htbp]
\centering
\includegraphics[width=0.45\textwidth]{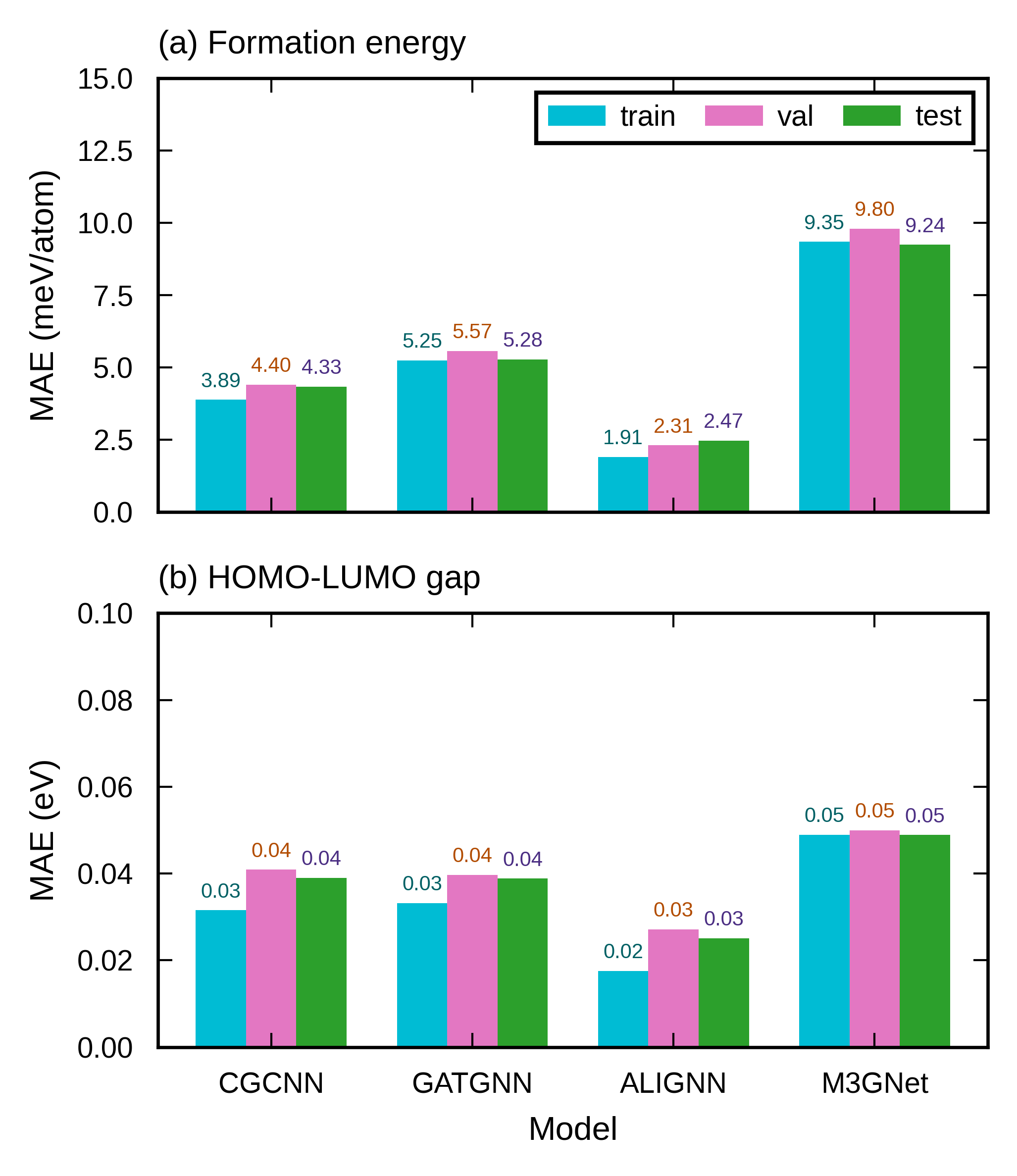}
\caption{Train, validation, and test MAE values for the ordered domain (single and double perovskite). (a) Formation-energy error (meV/atom). (b) HOMO-LUMO gap errors (eV).}
\label{fig03:model-benchmark}
\end{figure}

To evaluate the performance of various trained GNN models, we use the converged best model to predict the formation energy and HOMO-LUMO gap for the training, validation, and test sets curated from ordered single and double perovskite dataset.
The prediction error was quantified using the mean absolute error (MAE) between the model prediction and the DFT reference value:

\begin{equation}
\mathrm{MAE} = \frac{1}{N}\sum_{i=1}^{N} \left| \hat{y}_i - y_i^{\mathrm{DFT}} \right|,
\end{equation}

where $N$ is the number of structures, $\hat{y}_i$ is the predicted value, and $y_i^{\mathrm{DFT}}$ is the corresponding DFT target value.
The MAE results in Fig.~\ref{fig03:model-benchmark} show that ALIGNN gives the best overall performance for both target properties; the corresponding root-mean-square error (RMSE) results are provided in Fig.~S1 in the SI.
For formation-energy prediction (Fig.~\ref{fig03:model-benchmark}a), ALIGNN achieves the lowest test MAE of 2.47~meV/atom.
CGCNN and GATGNN give larger but still moderate test errors of 4.34 and 5.28~meV/atom, respectively, while M3GNet gives the highest test error of 9.25~meV/atom.
The train, validation, and test bars remain close for all architectures, indicating that the benchmark is not dominated by severe overfitting.
The main exception is the larger absolute error of M3GNet, which is consistent with its weaker convergence in Fig.~\ref{fig02:training-history}.

\FloatBarrier

For HOMO-LUMO gap prediction (Fig.~\ref{fig03:model-benchmark}b), ALIGNN also gives the lowest test MAE of 0.03~eV.
CGCNN and GATGNN give nearly identical test errors of 0.04~eV, and M3GNet gives a slightly larger test error of 0.05~eV.
The gap between ALIGNN and the pairwise graph models is smaller for $E_g$ than for $E_f$, but the same qualitative trend remains: the angle-aware model is the most accurate on the held-out test set of ordered domain. Overall, these benchmark results identify ALIGNN as the most accurate architecture for the ordered-perovskite domain, while CGCNN remains a reliable and simpler baseline.

\begin{figure}[!htbp]
\centering
\includegraphics[width=0.95\textwidth]{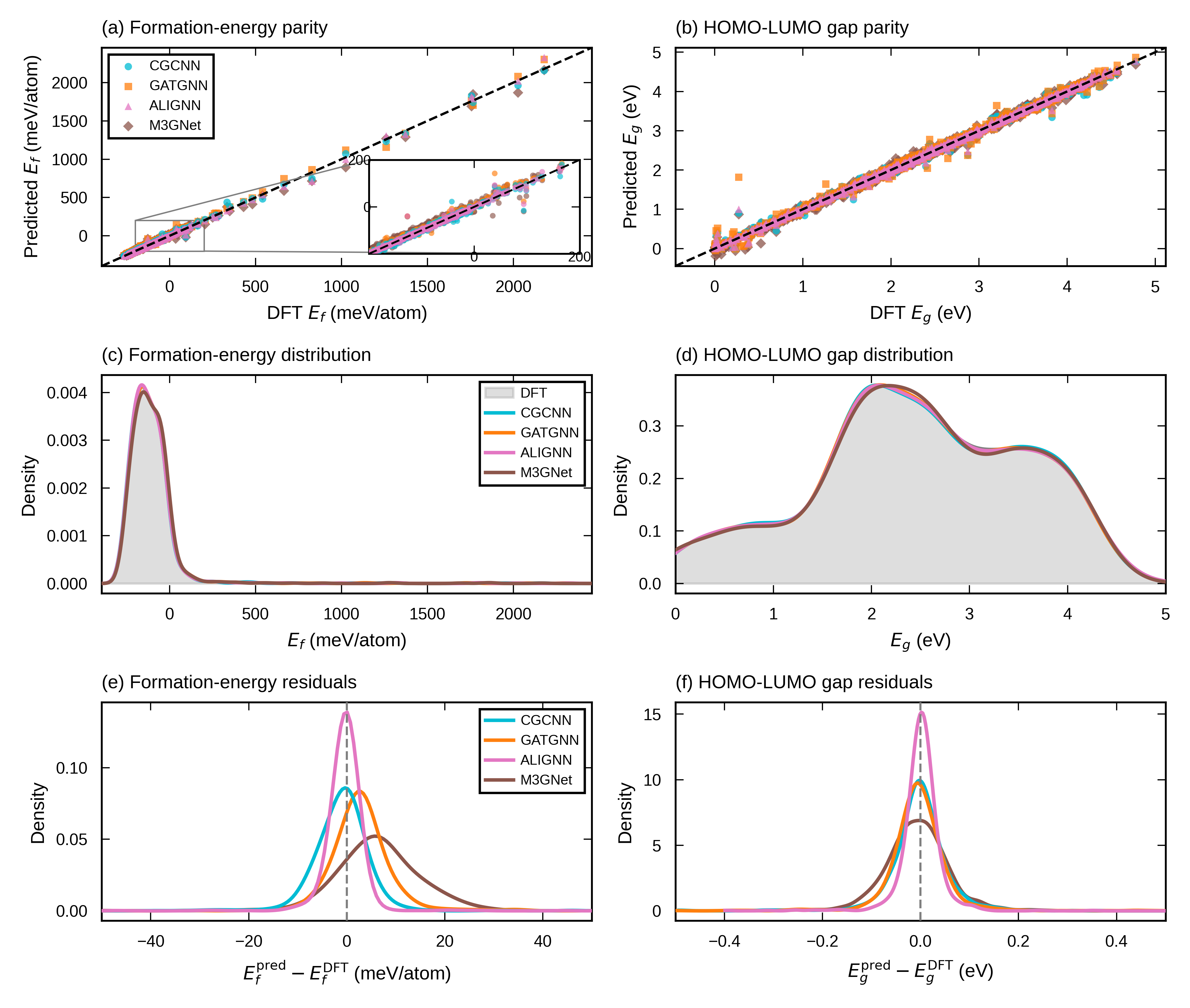}
\caption{Prediction quality of the trained GNN models on the ordered single and double perovskite test set.
Parity plots for (a) formation-energy and (b) HOMO-LUMO gap predictions.
Distribution of the calculated DFT and model predicted values of (c) formation energy and (d) HOMO-LUMO gap.
Residual distributions of all GNN models for (e) formation energy and (f) HOMO-LUMO gap.
}
\label{fig04:prediction-quality}
\end{figure}
\FloatBarrier

The parity plots in Fig.~\ref{fig04:prediction-quality}(a) and~\ref{fig04:prediction-quality}(b) show that the predicted values closely follow the DFT reference values for both targets.
For formation energy, the inset in Fig.~\ref{fig04:prediction-quality}(a) highlights the low-energy region, where many structures are concentrated and where small errors on the meV/atom scale are important for ranking similar candidates.
For HOMO-LUMO gap, the models reproduce the DFT values across the full 0--5~eV range, although a few outlier with low-gap and intermediate-gap structures deviate from the ideal parity line.

The distribution plots in Fig.~\ref{fig04:prediction-quality}(c) and~\ref{fig04:prediction-quality}(d) show that the predicted target distributions has a significant overlap with the DFT distribution.
This indicates that the models reproduce not only individual test values but also the overall property distributions of the ordered test set.
The residual distributions in Fig.\ref{fig04:prediction-quality}(e) and~\ref{fig04:prediction-quality}(f) compare all the trained GNN models for both the formation-energy and HOMO-LUMO gap prediction, respectively.
For both targets, the residuals are approximately symmetric and centered near zero, indicating that neither model exhibits a strong systematic prediction bias on the ordered test set. The distribution of ALIGNN model are much narrower than the corresponding CGCNN distributions, consistent with the lower ALIGNN test MAEs for both targets. Comparatively, other GNN models GATGNN and M3GNET have distributions which are asymmetric, i.e. shifted towards higher formation energy and slightly shifted towards lower HOMO-LUMO gap. This distribution of GATGNN is comparable to that of CGCNN. And, M3GNET displays largest spread in their distribution among the four models. 

Analysis of the trends on the distributions and error residue on the ordered-domain leads to the conclusion that the models learn a coherent structure--property relationship rather than only matching average target distributions.
ALIGNN provides the most accurate and least dispersed predictions, and this makes it the appropriate model for the subsequent ordered-to-HEPO transfer analysis. In addition, we also conduct analysis on the CGCNN model as a representative of GNN model devoid of three-body term and a suitable baseline reference. 

\subsection*{Error analysis across chemical species}

\begin{figure}[!htbp]
\centering
\includegraphics[width=0.95\textwidth]{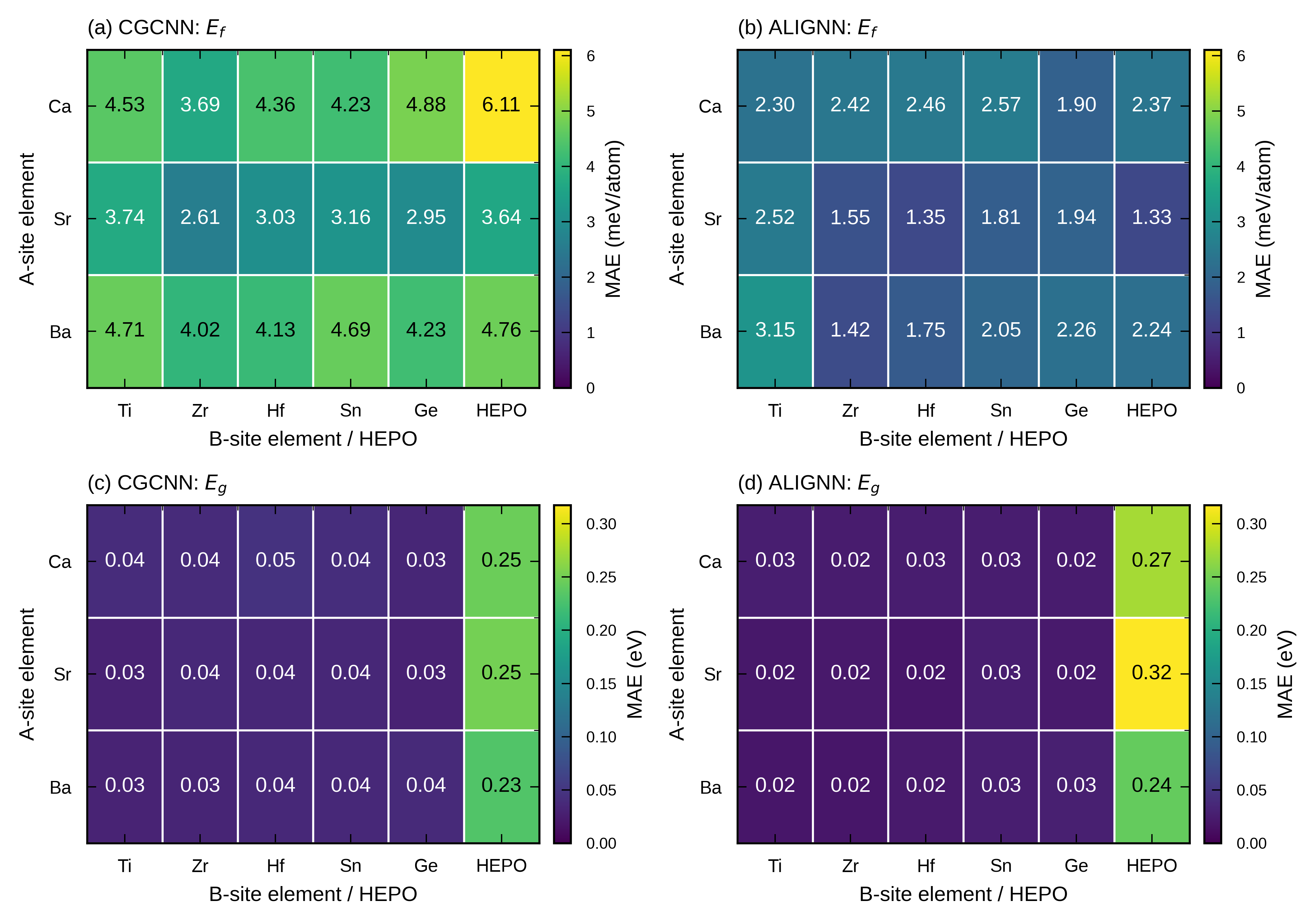}
\caption{Chemistry-resolved prediction errors on held out test set of ordered and disordered domain. CGCNN MAE maps for (a) formation energy $E_f$ (meV/atom) and (b) HOMO-LUMO gap $E_g$ (eV).
ALIGNN MAE maps for (c) formation energy $E_f$  and (d) HOMO-LUMO gap $E_g$ .
Rows denote the A-site element, and columns denote the ordered B-site element or the HEPO transfer group.}
\label{fig05:chemistry-error}
\end{figure}
\FloatBarrier

We now focus on a chemistry-based error analysis to determine whether the ordered domain (single and double perovskites) errors remain uniform across different chemistry groups and identify where the transfer errors of the GNN models trained on ordered domain becomes larger in disordered domain (HEPOs). As described in Sec. \textit{Dataset construction}, the HEPO dataset comprises three compositions with an identical set of five B-site elements [Ti, Zr, Hf, Sn, Ge], differing only in the A-site cation, which is either Ca, Sr, or Ba.
This systematic comparison allows us to evaluate whether the global test MAE provides a representative measure of performance across the chemical space. Therefore, we calculated the test-set MAE separately for each A-site and B-site combination in the form of heatmaps, as shown in Fig.~\ref{fig05:chemistry-error}; the corresponding RMSE results are provided in Fig.~S2 in the SI. Each heatmap cell represents the MAE for structures containing the corresponding A-site and B-site elements.
For example, the Ca--Ti cell reports the model error for test structures in the Ca--Ti chemistry group, such as CaTiO$_3$ and Ca$_2$TiB'O$_6$-related ordered perovskite configurations containing Ca on the A site and Ti on the B site.
The final HEPO column in each heatmap is treated separately because it corresponds to chemically disordered SQS structures rather than to a single ordered B-site element.

For formation-energy prediction, CGCNN gives errors between 2.61 and 4.88~meV/atom for ordered perovskites.
The lowest CGCNN error occurs for Sr--Zr, while the larger errors occur for Ca--Ge, Ba--Ti, and Ba--Sn.
Across the HEPO column, the errors in formation energy for all the three A-site cations Ca, Sr, Ba
are within the same scale with values of 6.11, 3.64, and 4.76~meV/atom, respectively. This illustrates that formation-energy prediction is relatively robust when moving from ordered B-site chemistry to the HEPO transfer group.

ALIGNN reduces the formation-energy errors nearly across all ordered chemistry groups. The ordered-chemistry MAE values mostly lie between 1.42 to 3.15~meV/atom, with particularly low errors for Sr--Hf and Ba--Zr. The HEPO formation-energy errors are also low for ALIGNN, giving 2.37, 1.33, and 2.24~meV/atom for the Ca-, Sr-, and Ba-based HEPO groups, respectively. Thus, the ALIGNN MAE remains below 3.15~meV/atom across the ordered group and below 2.37~meV/atom across the three HEPO families.

The situation is completely different for HOMO-LUMO gap predictions where errors on the HEPO systems are systematically and drastically higher than on the ordered domain, independent of the considered A-site element, for both CGCNN and ALIGNN models. CGCNN gives ordered HOMO-LUMO gap MAE values of 0.03--0.05~eV, whereas the HEPO column increases to 0.23--0.25~eV. ALIGNN reduces the ordered-domain HOMO-LUMO gap errors to 0.02--0.03~eV, but the HEPO column remains much larger, with values of 0.24--0.32~eV. In general, the trends are similar for both the CGCNN and ALIGNN models. Thus, the challenge in training HOMO-LUMO gap is not necessarily the interpolation among ordered chemistries, but the transfer from ordered local environments to chemically disordered local environments in HEPO.

Beyond the models trained in this work, we further evaluated existing pretrained CGCNN and ALIGNN models for HOMO–LUMO gap prediction using their available PBE checkpoints~\cite{yan2022periodic}. Pretrained formation-energy models were excluded from this analysis because their reference states are not directly compatible with the oxide-referenced definition adopted here. The chemistry-resolved MAEs on HOMO-LUMO gap range from 0.36 to 1.17~eV for CGCNN and from 0.32 to 0.80~eV for ALIGNN. Although, pretrained ALIGNN performs better for several chemistries, both models are substantially less accurate than those trained on the ordered perovskite dataset, indicating that perovskite-specific training or fine-tuning is required for quantitative HEPO screening. The corresponding heatmaps are provided in Fig.~S3 in the SI.

Overall, the chemistry-resolved analysis displays two different transfer behaviors.
Formation energy is consistent across ordered and HEPO chemistry groups, especially for ALIGNN, while HOMO-LUMO gap prediction is accurate within ordered perovskites but much more sensitive to chemical disorder in HEPOs. 
The foundation-model comparison demonstrates that pretrained graph models require specific adaptation to the considered dataset chemistry to reach the accuracy needed for HOMO-LUMO gap screening. 

\subsection*{Feature embedding and UMAP analysis}

\begin{figure}[!htbp]
\centering
\includegraphics[width=0.95\textwidth]{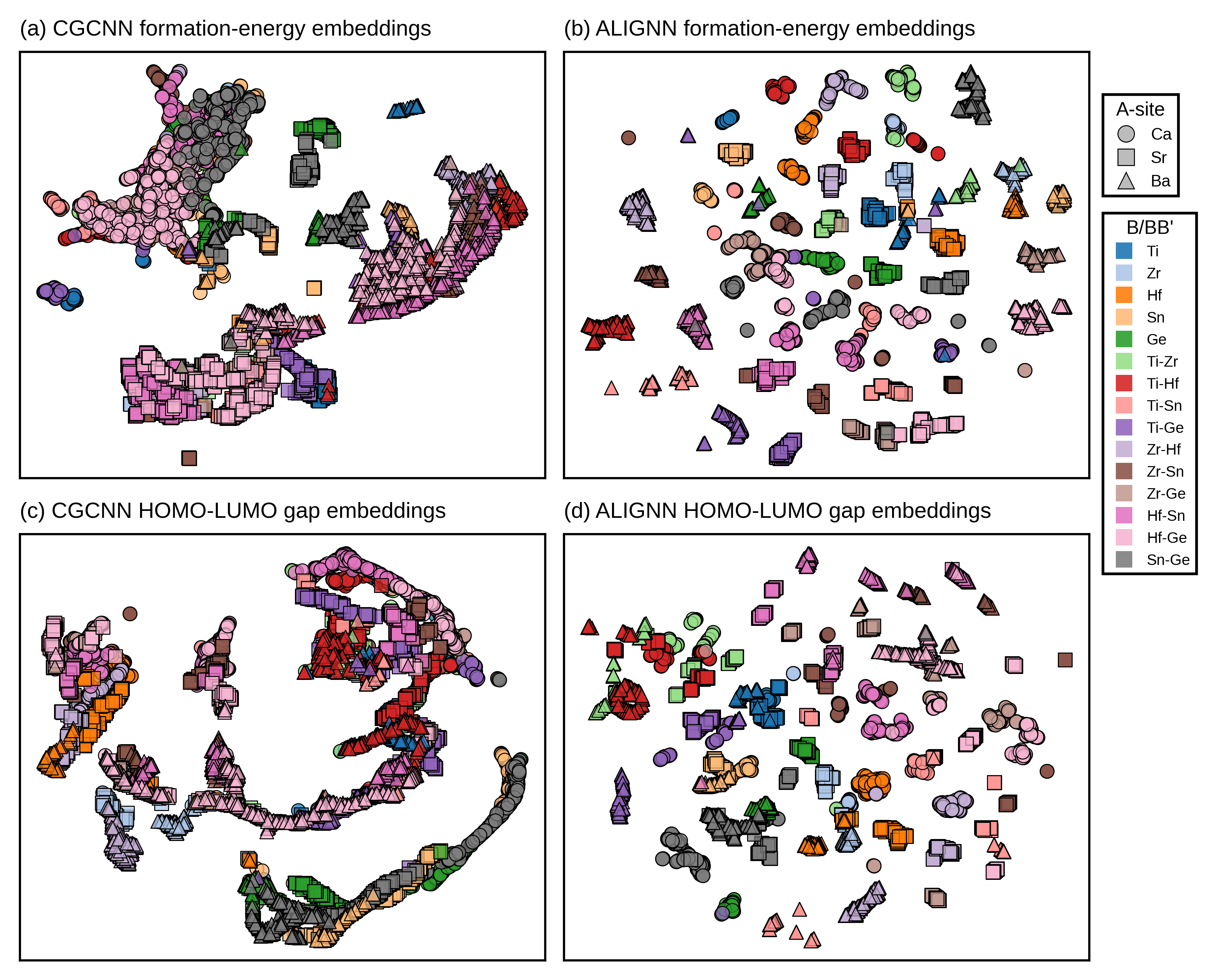}
\caption{UMAP projections of graph-level embeddings for
(a) CGCNN formation-energy prediction,
(b) ALIGNN formation-energy prediction,
(c) CGCNN HOMO-LUMO gap prediction, and
(d) ALIGNN HOMO-LUMO gap prediction.
Marker shape denotes the A-site element and color denotes the B-site or B/B$'$ group.}
\label{fig06:umap1}
\end{figure}

To inspect whether the trained graph models learned chemically and physically meaningful representations, graph-level embeddings were extracted from the trained models and projected using 
uniform manifold approximation and projection (UMAP) technique as discussed in Ref.~\cite{UMAP}. Before UMAP projection, the high-dimensional embeddings were standardized and compressed using principal component analysis (PCA)~\cite{PCA} to reduce noise and improve the stability of the two-dimensional visualization, as shown in Fig.~\ref{fig06:umap1}. CGCNN and ALIGNN were selected for this analysis because they represent two contrasting graph-learning descriptions. While, CGCNN is a simpler baseline model based mainly on atom--bond pair information, ALIGNN includes explicit bond-angle information and gives the best overall prediction accuracy. To confirm that the main clustering patterns are reproducible, UMAP projections of the same CGCNN and ALIGNN formation-energy embeddings generated using different random seeds are shown in Fig.~S4 in the SI.

\FloatBarrier

For formation-energy prediction, the CGCNN embedding in Fig.~\ref{fig06:umap1}(a) forms broad, connected manifolds with overlap between different B-site and B/B$'$ groups. Some A-site separation is visible from the marker shapes, but many composition groups remain mixed. This indicates that CGCNN captures general chemical trends but does not fully separate the ordered-perovskite chemistry space. In contrast, the ALIGNN formation-energy embedding in Fig.~\ref{fig06:umap1}(b) is divided into many compact clusters. Many of these islands are dominated by specific B-site or B/B$'$ groups and also show A-site-dependent structure. This stronger separation is consistent with the lower ALIGNN formation-energy MAE and suggests that angle-aware message passing helps organize stability-related structural motifs.

For HOMO-LUMO gap prediction, the CGCNN embedding in Fig.~\ref{fig06:umap1}(c) again forms extended continuous manifolds. Several composition groups occupy local regions, but the latent space remains relatively mixed, especially among the B/B' groups of the double perovskites. The ALIGNN HOMO-LUMO gap embedding in Fig.~\ref{fig06:umap1}(d) shows clearer local clustering and greater separation between composition groups. However, considerable overlap remains, which may reflect the combined effects of chemical composition and structural distortion on the HOMO-LUMO gap. Consistent with the prediction results, ALIGNN performs best for $E_g$, although its improvement over CGCNN and GATGNN is less pronounced than that observed for $E_f$.

The embeddings were also colored by their DFT target values to examine whether the learned representations organize structures by formation energy and HOMO-LUMO gap (Fig.~S5 in the SI).
The local color gradients show that nearby structures often have similar target values, while the A-site-dependent marker patterns indicate that the representations retain chemical information.

\begin{figure}[!htbp]
\centering
\includegraphics[width=0.95\textwidth]{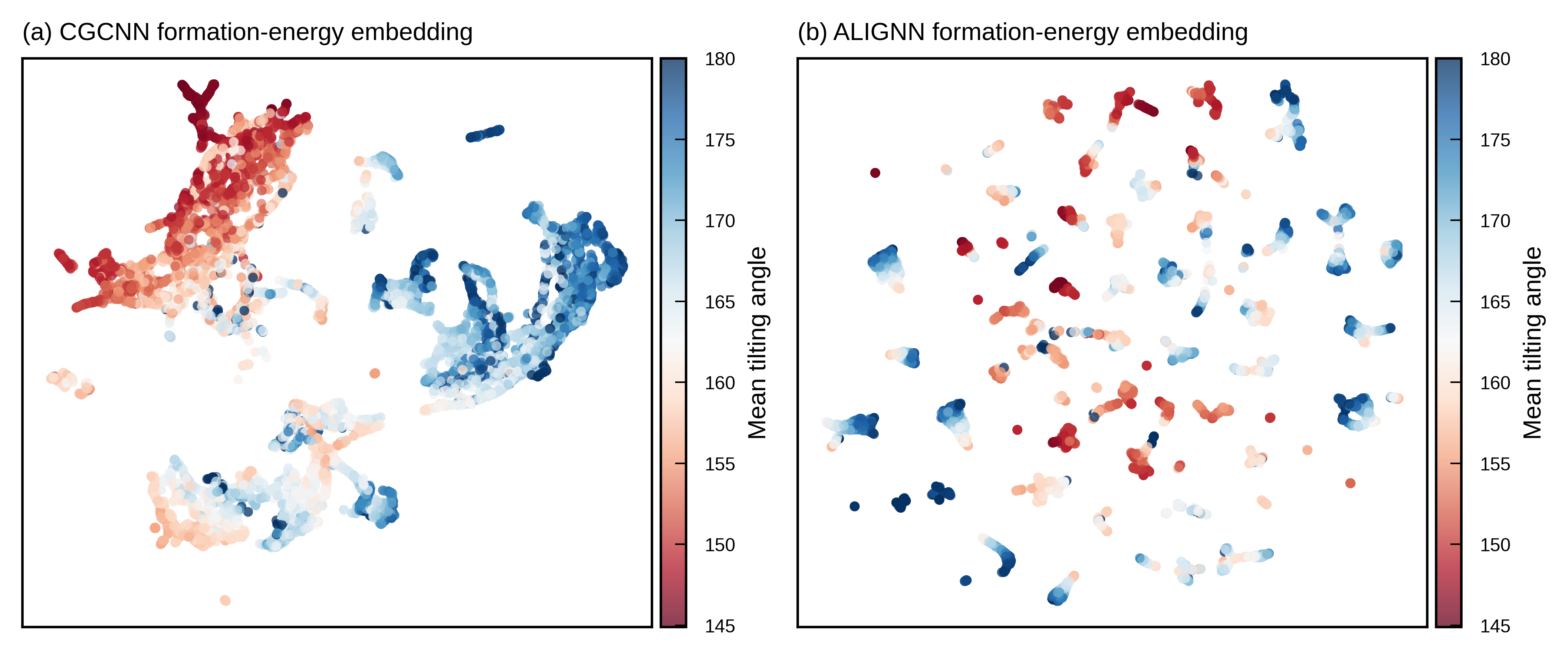}
\caption{UMAP projections of graph-level embeddings for
(a) CGCNN formation-energy prediction and
(b) ALIGNN formation-energy prediction.
Color denotes the average tilting angle in each structure.}
\label{fig07:umap2}
\end{figure}

Overall, the chemically colored UMAP results show that the graph embeddings encode composition and A-site identity, with ALIGNN giving clearer local separation than CGCNN. The remaining question is whether these learned embeddings also contain information about octahedral tilting, a structural descriptor that is central to perovskite distortions~\cite{glazer1972classification,woodward1997octahedral,xiang2017rules}. The color code in Fig.~\ref{fig07:umap2} represents formation-energy embedding colored by the average B--O--B tilting angle connecting neighboring BO$_6$ octahedra. An angle of $180^\circ$ indicates no tilting or distortion between neighboring BO$_6$ octahedra,
whereas values below $180^\circ$ indicate increasing octahedral tilting or distortion.

\FloatBarrier

The embedding of CGCNN model in Fig.~\ref{fig07:umap2}(a) contains broad regions with similar color, showing that tilting information is partly encoded even in the pairwise atom--bond graph representation.
However, the color gradient is distributed across large connected manifolds, indicating that CGCNN mixes tilting information with composition and A-site chemistry. In contrast, the projection for ALIGNN as shown in Fig.~\ref{fig07:umap2}(b) contains more compact local groups, and many of these groups have internally consistent tilting-angle values. This indicates that the angle-aware representation can organize structures according to both composition and local angular distortion. Nevertheless, the color distribution is not perfectly separated. This is expected as formation energy also depends on bond lengths, cation chemistry, and the strain associated with the relaxed structure.

Overall, the embedding analysis shows that the trained graph models learn chemically meaningful latent spaces and that ALIGNN provides a more resolved organization of both composition and tilting-related structural variation. This analysis thus provides supporting evidence that the best predictive accuracy is obtained when the graph representation explicitly includes angular structural information (three-body term).

\subsection*{Transferability from ordered perovskites to HEPOs}

The HEPO structures and their DFT properties were generated in our previous work~\cite{jomaa2026aiida,untarabut2026entropy} and combined here with the ordered-perovskite data to form a unified database. The HEPO dataset contains Ca-, Sr-, and Ba-based compositions. For each A-site family, the five B-site elements [Ti, Zr, Hf, Sn, Ge] were combined in an equimolar concerntration and their chemical disorder is represented via various SQS structural models. This dataset was used as an external test to probe whether the structure--property relationships learned from ordered perovskites transfer to chemically disordered HEPOs and to compare prediction errors across the three A-site families.

\begin{figure*}[!htbp]
\centering
\includegraphics[width=0.95\textwidth]{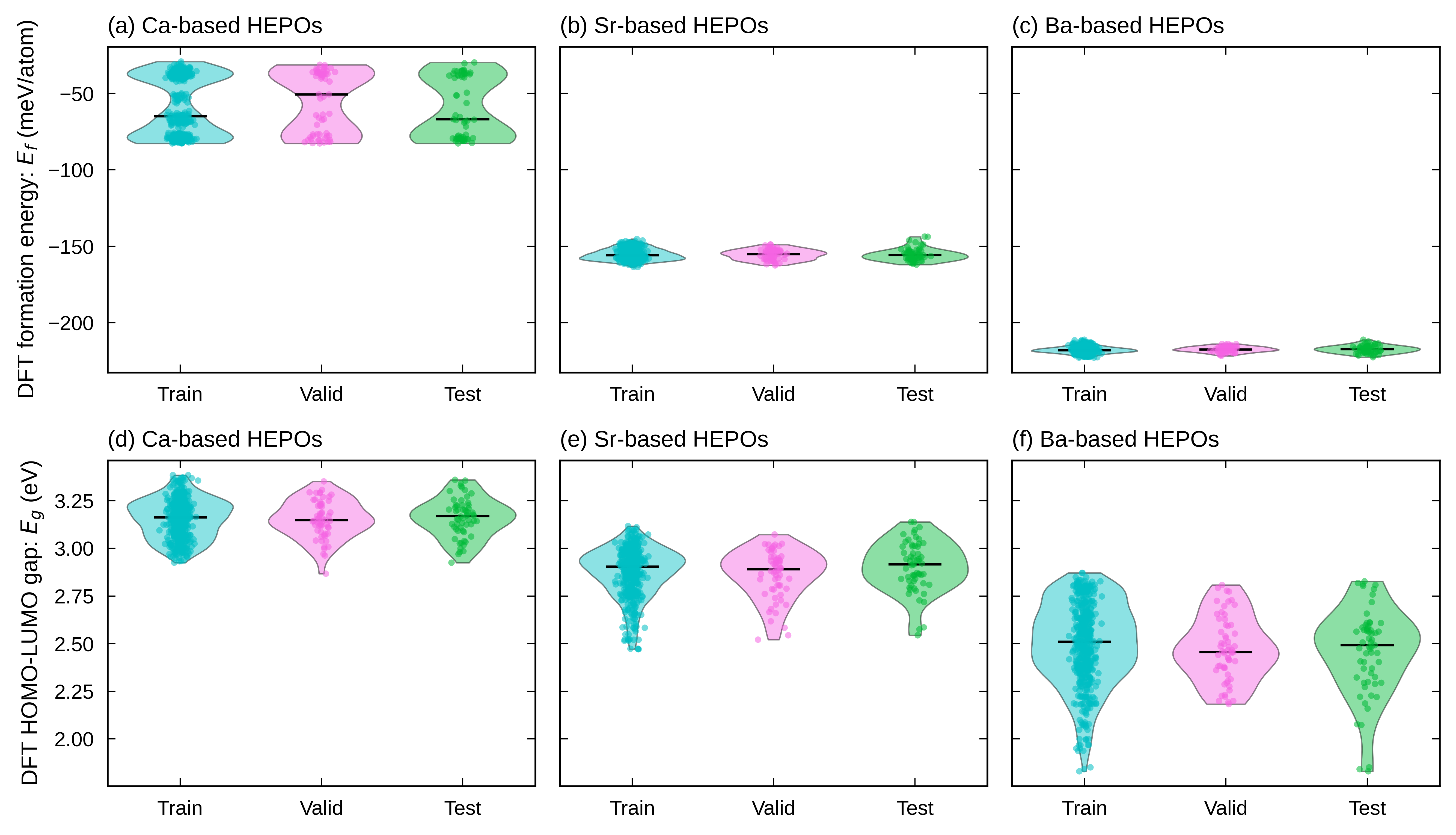}
\caption{HEPO target-property distributions across the training, validation, and test subsets.
(a--c) DFT formation-energy distributions and
(d--f) DFT HOMO--LUMO-gap distributions for the Ca-, Sr-, and Ba-based HEPO families, respectively.
Cyan, pink, and green violins represent the training, validation, and test subsets, while the overlaid points denote individual structures.
The black horizontal line in each violin indicates the mean value.}
\label{fig08:hepo_dataset}
\end{figure*}
\FloatBarrier

The distribution of HEPO dataset across training, validation and test-set are shown in Fig.~\ref{fig08:hepo_dataset}, where
Fig.~\ref{fig08:hepo_dataset}(a)--(c) show the formation-energy distribution and
Fig.~\ref{fig08:hepo_dataset}(d)--(f) show the HOMO--LUMO-gap distributions, respectively, for the Ca-, Sr-, and Ba-based HEPO families. While, the oxide-referenced formation energy becomes more negative from Ca- to Sr- to Ba-, the average HOMO-LUMO gap decreases across these systems.
Within the reference scheme used here, The Ba-based HEPOs have the most negative formation energies relative to the selected oxide reference phases. As summarized in Table~\ref{tab:hepo_target_statistics}, the training, validation, and test subsets have similar means and standard deviations within each A-site family, indicating comparable target-property distributions across the three subsets.

\begin{table*}[!htbp]
\centering
\caption{Average and standard deviation of the HEPO target properties for the training, validation, and test subsets.}
\label{tab:hepo_target_statistics}
\begin{tabular}{llrrrr}
\toprule
A-site & Subset & \multicolumn{2}{c}{Formation energy (meV/atom)} &  \multicolumn{2}{c}{HOMO--LUMO gap (eV)}\\
\midrule 
       &        & Average & Standard & Average & Standard \\
       &        &         & deviation &        & deviation \\
\midrule
\midrule
Ca & Train & $-58.50$  & $19.36$ & $3.16$ & $0.10$ \\
Ca & Valid & $-55.16$  & $19.98$ & $3.15$ & $0.10$ \\
Ca & Test  & $-59.43$  & $19.84$ & $3.16$ & $0.10$ \\
\addlinespace
Sr & Train & $-155.30$ & $3.65$ & $2.88$ & $0.13$ \\
Sr & Valid & $-155.44$ & $3.47$ & $2.86$ & $0.13$ \\
Sr & Test  & $-155.39$ & $4.10$ & $2.91$ & $0.13$ \\
\addlinespace
Ba & Train & $-217.72$ & $2.17$ & $2.50$ & $0.22$ \\
Ba & Valid & $-217.30$ & $1.86$ & $2.47$ & $0.17$ \\
Ba & Test  & $-217.25$ & $2.44$ & $2.46$ & $0.23$ \\
\bottomrule
\end{tabular}
\end{table*}

For the strict transfer test, we selected the ALIGNN model trained only on the ordered-perovskite dataset (as discussed in the previous section) and used it to predict formation energies and HOMO-LUMO gaps of the HEPO structures from the held-out test set. This is a blind-transfer test because the model did not use any HEPO structures during training or validation. The comparison therefore probes whether the structural and chemical patterns learned from ordered perovskites remain useful for high-entropy perovskite oxides.

\begin{figure}[!htbp]
\centering
\includegraphics[width=0.95\textwidth]{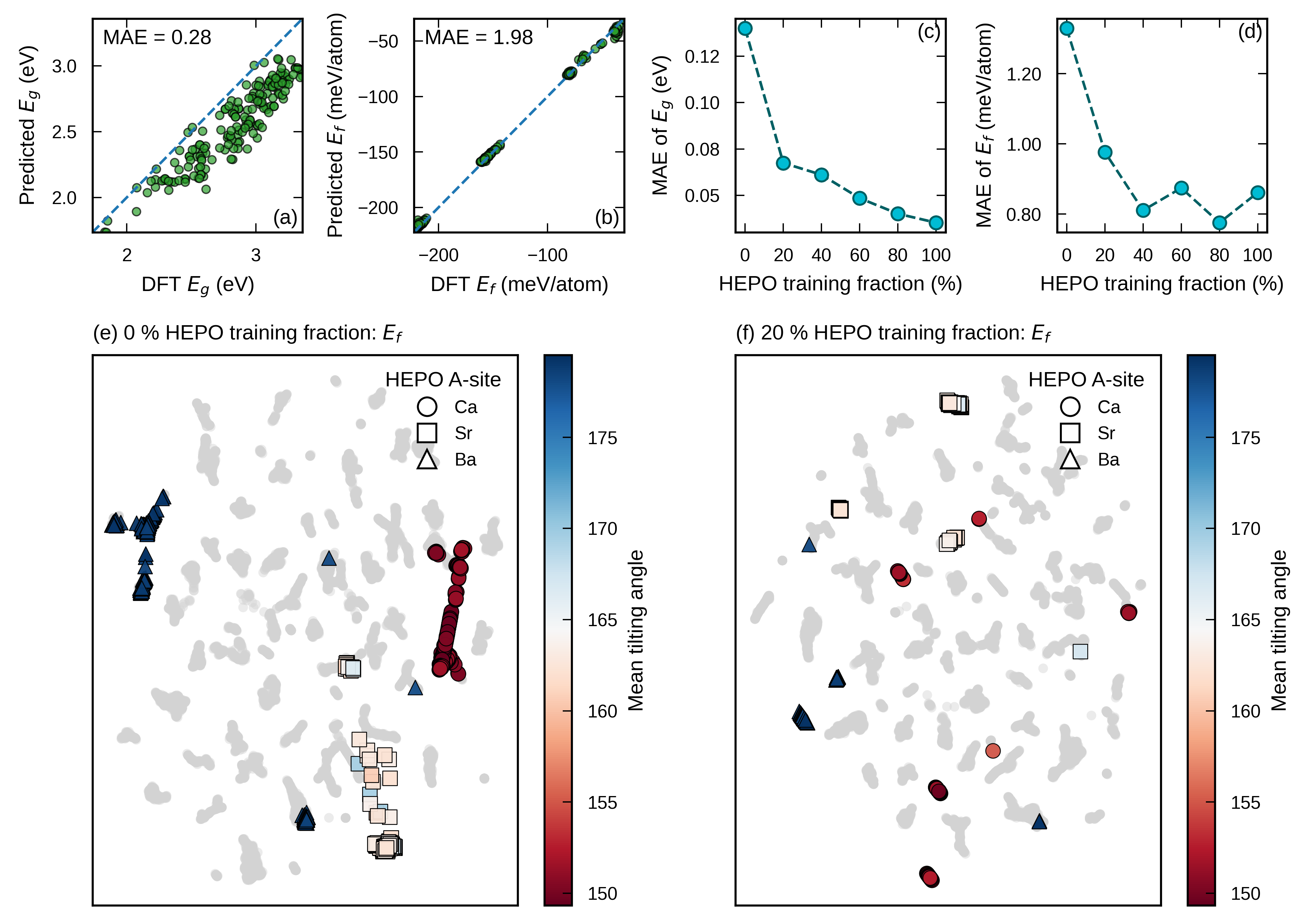}
\caption{Ordered perovskite to HEPO transferability analysis.
Blind-test parity plots comparing DFT reference values and ALIGNN predictions for (a) HOMO-LUMO gap (eV) and (b) formation energy (meV/atom). HEPO fine-tuning and data-efficiency analysis showing the MAE as a function of the HEPO training fraction for (c) HOMO-LUMO gap (eV) and (d) formation energy (meV/atom).
UMAP projections of formation-energy embeddings for (e) 0\% and (f) 20\% HEPO training fractions, with HEPO structures marked by A-site family and colored by mean tilting angle.
}
\label{fig09:hepo}
\end{figure}

The transfer results are shown in Fig.~\ref{fig09:hepo}(a) and~\ref{fig09:hepo}(b) as parity plots between the DFT reference and predicted values for HOMO-LUMO gap and formation energy, respectively.
The MAE values computed on the  HEPO test subset are 0.28~eV for HOMO-LUMO gap prediction and 1.98~meV/atom for formation-energy prediction. The formation-energy error is comparable to the ordered-perovskite test error of 2.47~meV/atom, indicating that the stability-related structural patterns learned from ordered perovskites transfer well to the HEPO structures. In contrast, the HOMO-LUMO gap error is much higher than the ordered-perovskite test error of 0.03~eV. 

HOMO-LUMO gaps tend to be systematically under predicted for HEPO structures and the chemical disorder on B-sites have a global effect on the HOMO-LUMO gap that the model cannot capture. This indicates that electronic-structure prediction is more sensitive than formation-energy prediction to chemically disordered local environments in HEPOs.

\FloatBarrier

To analyze the data efficiency of HEPO fine-tuning, the HEPO training subset was divided into fractions of 0.0, 0.2, 0.4, 0.6, 0.8, and 1.0, where 0.0 corresponds to using no HEPO structures in training and 1.0 corresponds to using the full HEPO training subset. For each fraction, the selected HEPO training structures were combined with the ordered-perovskite training dataset, and a new ALIGNN model was trained from scratch. The HEPO validation subset was used for model validation, and the HEPO test subset was used only for the final accuracy evaluation. The resulting MAE values are shown in Fig.~\ref{fig09:hepo}(c) and~\ref{fig09:hepo}(d) for HOMO-LUMO gap and formation-energy prediction, respectively.

The 0.0 HEPO training fraction is an important reference point because the model is still trained only on the ordered-perovskite dataset. However, unlike the fully blind-transfer test, the HEPO validation subset is used for model selection. For HOMO-LUMO gap prediction, this reduces the HEPO test MAE from 0.28~eV in the blind-transfer test to 0.14~eV at the 0.0 HEPO training fraction. This improvement shows that part of the blind-test error comes from a validation-domain mismatch: a model selected using only ordered-perovskite validation data is not necessarily the best checkpoint for HEPO prediction. Using a HEPO validation set helps select a model that is better matched to the HEPO distribution, even before any HEPO structures are added to the training set. Therefore, the 0.0-fraction result should be interpreted as ordered-domain training with HEPO-guided model selection rather than as a fully blind-transfer result.

Adding HEPO structures to the training set further reduces the HOMO-LUMO gap MAE\@.
The largest improvement occurs when the HEPO training fraction increases from 0.0 to 0.2, where the MAE decreases rapidly from 0.14~eV to 0.07~eV. As the HEPO training subset corresponds to 80\% of the full HEPO dataset, a 0.2 training fraction uses only 16\% of all HEPO structures for training.
The MAE then decreases consistently to 0.06, 0.05, $\sim$0.045, and 0.04~eV at HEPO training fractions of 0.4, 0.6, 0.8, and 1.0, respectively. This indicates that a relatively small HEPO training fraction is sufficient to produce the largest adaptation to the HEPO HOMO-LUMO gap distribution, while additional HEPO data provide smaller but systematic improvements. At the full HEPO training fraction, the HOMO-LUMO gap MAE is 0.04~eV, representing a reduction of approximately 87\% relative to the blind-transfer MAE\@.

For formation-energy prediction, the MAE is already low for the 0.0 HEPO training fraction and remains below the blind-transfer error after HEPO-guided model selection. The formation-energy MAEs are 1.33, 0.98, 0.81, 0.87, 0.78, and 0.86~meV/atom at HEPO training fractions of 0.0, 0.2, 0.4, 0.6, 0.8, and 1.0, respectively. The minimum MAE of 0.78~meV/atom is obtained with a HEPO training fraction of 0.8, corresponding to a reduction of approximately 61\% relative to the blind-transfer MAE of 1.98~meV/atom. Unlike the HOMO-LUMO gap results, the formation-energy error is not strictly monotonic with the amount of HEPO training data, as small increases occur at fractions of 0.6 and 1.0. This weaker and non-monotonic dependence on HEPO fraction suggests that formation energy transfers more directly from ordered perovskites to HEPOs, whereas HOMO-LUMO gap prediction benefits more strongly from even a small amount of HEPO-specific training data. The complete blind-transfer and fine-tuning results are summarized in Table~S4 in the SI.

The UMAP panels in Fig.~\ref{fig09:hepo}(e) and~\ref{fig09:hepo}(f) provide a representation-level view of this adaptation.
At 0\% HEPO training, the HEPO structures already occupy A-site dependent regions, indicating that the ordered-domain model recognizes the chemical differences among the Ca-, Sr-, and Ba-based families.
However, these regions are less compact and less clearly separated because the model has not encountered chemically disordered HEPO environments during training.
After adding 20\% of the HEPO training subset, the three A-site families form more compact and distinct clusters, showing that a relatively small amount of HEPO-specific data improves the organization of the disordered structures in the latent space.
Within these composition-dependent clusters, the continuous color variation with mean tilting angle reveals additional organization by structural distortion.
The simultaneous separation by A-site family and variation with tilting angle indicate that ALIGNN encodes both chemical composition and the local structural disorder associated with octahedral tilting.
These results support the interpretation that fine-tuning with the limited HEPO dataset adapts the learned representation to the disordered domain while retaining chemically and structurally meaningful relationships.

Additional projections colored by domain, A-site element, and mean tilting angle are provided in Fig.~S6 in the SI. They further show that the limited HEPO fine-tuning integrates the disordered structures into the learned representation while preserving chemical and structural organization.

Overall, the HEPO transfer results show that the ordered-perovskite training is already sufficient for formation-energy transfer but not for a reliable HOMO-LUMO gap prediction.
For screening HEPO stability, the ordered model is therefore a useful transferable predictor.
For HEPO electronic-structure screening, however, the model should include at least a small HEPO-specific training fraction to correct the ordered-to-disordered domain shift.

\section*{Conclusions}
In this work, we evaluated ordered-to-disordered transfer learning using a DFT dataset that connects single ABO$_3$ and double A$_2$BB'O$_6$ perovskite structures to high entropy perovskite structures.
By keeping the A-site and B-site element pools fixed while moving from ordered to chemically disordered local environments, the dataset separates ordinary interpolation within ordered perovskites from the more demanding question of transfer to HEPOs.

Among the tested four graph neural networks, ALIGNN provided the most accurate and consistent predictions for both formation energy and HOMO-LUMO gap. Within the single and double perovskite ordered test set, ALIGNN achieved the lowest MAEs of  2.47~meV/atom for formation energy and 0.03~eV for HOMO-LUMO gap. Its prediction performance was found to be better than that of CGCNN, GATGNN, and M3GNet; which may be related to its direct use of bond-angle information in the graph message construction. This is particularly important for perovskite oxides where the angular distortions are known to strongly affect both structural stability and electronic properties. In addition, the UMAP analysis on model embedding also supports this result, i.e. the ALIGNN representation forms more compact regions associated with chemical composition with a clear sensitivity to structural distortion within individual clusters.

The HEPO transfer results show that formation energy and HOMO-LUMO gap respond differently to the transition from ordered to disordered structures. Formation-energy prediction transfers well across both these domains. The ALIGNN model trained only on ordered structures gave a blind HEPO MAE of 1.98~meV/atom, which is similar to its error on the ordered test set. This result suggests that the structural and chemical features controlling stability in ordered perovskites are also useful for predicting the stability of HEPO structures.

In contrast, HOMO–LUMO gap prediction is more strongly influenced by chemical disorder. This is evident from the performance of GNN models trained exclusively on the ordered domain, which exhibit an approximately 10-fold higher error on the HEPO blind test set, accompanied by a systematic underestimation of the HOMO-LUMO gaps. However, this limitation can be substantially mitigated by incorporating a small amount of HEPO-specific data: adding only 20\% of the HEPO training subset reduces the MAE to 0.07 eV, while using the full HEPO training subset further decreases it to 0.04 eV. In comparison, the improvement in formation-energy prediction is minimal with increasing amounts of HEPO training data. These results indicate that formation energies transfer more readily from ordered perovskites to HEPOs, whereas accurate HOMO–LUMO gap prediction requires the inclusion of HEPO-specific training data.

Through our investigation of different GNN models on curated perovskite oxide dataset we conclude that while the ordered perovskites are an effective training dataset for screening HEPO stability, HEPO HOMO-LUMO gap screening can greatly benefits from a small amount of HEPO-specific calibration. A practical strategy is therefore to first train graph neural networks using ordered-perovskite calculations, use angle-aware models such as ALIGNN, and then add a small, targeted HEPO dataset when electronic properties are the primary prediction targets. This approach reduces the number of expensive HEPO DFT calculations while maintaining a clear physical connection between ordered-perovskite structures, local chemical disorder, and material-property prediction.

\section*{Methods}
\label{sec2:Method}

\subsection*{Dataset construction}

The dataset consists of 10898 relaxed ordered ABO$_3$ and A$_2$BB'O$_6$ perovskites and 1810 relaxed HEPO structures.
The set of A-site elements was restricted to [Ca, Sr, Ba] and the B-site elements were restricted to
[Ti, Zr, Hf, Sn, Ge] for this study. The chemical and structural disorder in three equimolar HEPO families were represented by special quasirandom structures (SQSs) constructed from the same elemental pool~\cite{SQS}. Composition-resolved counts, the SQS compositions, and the data splits are provided in Table~S1 in Section~S1 of the SI.

DFT calculations were performed with Quantum ESPRESSO-7.2~\cite{hohenberg1964inhomogeneous,kohn1965self,giannozzi2009quantum,giannozzi2017advanced}, using the PBE exchange--correlation functional~\cite{perdew1996generalized}. The chemical species were described by the optimized norm-conserving Vanderbilt pseudopotentials as availied in PseudoDojo webpage~\cite{hamann2013optimized,van2018pseudodojo,lejaeghere2016reproducibility}. A kinetic energy cut-off of 80-Ry plane-wave cutoff with grid spacing of $\sim$0.30 \AA$^-1$ for Monkhorst--Pack $k$-point sampling~\cite{monkhorst1976special} for the calculations.
The electronic HOMO-LUMO gap ($E_g$) and oxide-referenced formation energy ($E_f$) were used as the regression targets.
The reference phases are listed in Table~S2 
, and the normalization conventions and formation-energy equations are given in Section~S2 of the SI.

\subsection*{Machine-learning models}
CGCNN, GATGNN, ALIGNN, and M3GNet were trained as separate regressors for $E_f$ and $E_g$ using the same structural data splits.
All models used the same elemental feature initialization, while their graph constructions retained their architecture-specific distance, angle, attention, or three-body representations.
Architecture settings, cutoffs, optimization parameters, target standardization, early stopping, and checkpoint selection are reported in Table~S3 and Section~S3 of the SI.

\section*{Acknowledgments}

The authors acknowledge funding from the PEPR-DIADEM AMADEUS Project (ANR-23-PEXD-0003), funded by the French ANR\@.
S.S.\ acknowledges funding support from the French ANR associated with the Chair Junior Professor fund (ANR-22-CPJ1-0039-01). 
Calculations were performed by using resources from Grand Equipement National de Calcul Intensif (GENCI, grants no. A0160915123, A0180915123, A0200915123, A0200913426 and A0200910832). In addition, computational resources provided by the computing facilities Mésocentre de Calcul Intensif Aquitain (MCIA) of the Université de Bordeaux and of the Université de Pau et des Pays de l’Adour were also used for the project.

\section*{Author contributions}
P.U. designed the computational workflow, generated and analyzed the data, trained the machine-learning models, and wrote the manuscript draft. N.J. and S.C. generated the HEPO dataset.
S.C., O.M., and S.B. contributed to data interpretation and discussion.
A.B. and S.S. supervised the project, contributed to the conceptual design, and revised the manuscript. All authors approved the final version.

\section*{Competing interests}

The authors declare no competing interests.

\section*{Additional information}

\bibliographystyle{naturemag}
\bibliography{references}

\end{document}